\documentclass[12pt]{article}
\usepackage{graphicx}
\usepackage{subfig}
\usepackage{glossaries}
\usepackage{makecell}
\usepackage{upgreek}
\usepackage{gensymb}
\usepackage{braket}
\usepackage{etoolbox}

\patchcmd{\thebibliography}
  {\settowidth}
  {\setlength{\parsep}{0.01pt}\setlength{\itemsep}{0pt plus 0.01pt}\settowidth}
  {}{}

\newcommand\pubnumber{SNSN-323-63}
\newcommand\pubdate{\today}

\def\lancaster{Department of Physics\\
Lancaster University, Bailrigg, Lancaster, United Kingdom}

\def\Title#1{\begin{center} {\Large #1 } \end{center}}
\def\Author#1{\begin{center}{ \sc #1} \end{center}}
\def\Address#1{\begin{center}{ \it #1} \end{center}}

\newcommand\pubblock{\rightline{\begin{tabular}{l} \pubnumber\\
         \pubdate  \end{tabular}}}
\newenvironment{Abstract}{\begin{quotation}  }{\end{quotation}}
\newenvironment{Presented}{\begin{quotation} \begin{center} 
             PRESENTED AT\end{center}\bigskip 
      \begin{center}\begin{large}}{\end{large}\end{center} \end{quotation}}

\def\beq{\begin{equation}}
\def\eeq#1{\label{#1}\end{equation}}
\def\eeqn{\end{equation}}

\def\beqa{\begin{eqnarray}}
\def\eeqa#1{\label{#1}\end{eqnarray}}
\def\eeqan{\end{eqnarray}}

\let\bar=\overbar

\def\Dslash{\not{\hbox{\kern-4pt $D$}}}
\def\dslash{\not{\hbox{\kern-2pt $\del$}}}

\def\msb{{\bar{\ssstyle M \kern -1pt S}}}

\newacronym{DUNE}{DUNE}{the Deep Underground Neutrino Experiment}
\newacronym{LArTPC}{LArTPC}{Liquid Argon Time Projection Chamber}
\newacronym{FD}{FD}{Far Detector}
\newacronym{LBNF}{LBNF}{the Long Baseline Neutrino Facility}
\newacronym{FHC}{FHC}{Forward Horn Current}
\newacronym{RHC}{RHC}{Reverse Horn Current}
\newacronym{CDR}{CDR}{Conceptual Design Report}
\newacronym{CC}{CC}{Charged Current}
\newacronym{NC}{NC}{Neutral Current}
\newacronym{MC}{MC}{Monte Carlo}
\newacronym{LBL}{LBL}{Long BaseLine}
\newacronym{CPV}{CPV}{Charge Parity Violation}
\newacronym{CP}{CP}{Charge Parity}
\newacronym{MH}{MH}{Mass Hierarchy}
\newacronym{MO}{MO}{Mass Ordering}
\newacronym{NO}{NMO}{Normal Mass Ordering}
\newacronym{IO}{IMO}{Inverted Mass Ordering}
\newacronym{GUT}{GUT}{Grand Unified Theories}
\newacronym{BSM}{BSM}{Beyond the Standard Model}
\newacronym{CNN}{CNN}{Convolutional Neutral Network}
\newacronym{LAr}{LAr}{Liquid Argon}
\newacronym{GAr}{GAr}{Gaseous Argon}
\newacronym{SP}{SP}{Single Phase}
\newacronym{DP}{DP}{Dual Phase}
\newacronym{ND}{ND}{Near Detector}
\newacronym{FNAL}{FNAL}{the Fermi National Accelerator Laboratory}
\newacronym{SURF}{SURF}{the Sanford Underground Research Facility}

\begin{document}
\begin{titlepage}
\pubblock

\vfill
\Title{DUNE: Status and Perspectives}
\vfill
\Author{ Dominic Brailsford, for the DUNE collaboration}
\Address{\lancaster}
\vfill
\begin{Abstract}
\noindent The Deep Underground Neutrino Experiment (DUNE) provides a rich science program with a focus on neutrino oscillations and other beyond the standard model physics.  The high-intensity, wide-band neutrino beam will be produced at the Fermi National Accelerator Laboratory (FNAL) and will be directed to the 40~kt liquid argon far detector at the Sanford Underground Research Facility, 1300~km from FNAL. The primary goals of the experiment are to determine the ordering of neutrino masses and to measure the CP violating phase, $\delta_{\textrm{CP}}$. The underground location of the large DUNE far detector and its excellent energy and spatial resolution will allow also for non-accelerator physics programs predicted by grand unified theories, such as nucleon decay or $n$---$\bar{n}$ oscillations.  Moreover, DUNE will be sensitive to the electron neutrino flux from a core-collapse supernova, providing valuable information on the phenomenon's underlying mechanisms. This ambitious project requires extensive prototyping and a testing program to guarantee that all parts of the technology are fully understood and well tested. Two such prototypes, in both single phase (ProtoDUNE-SP) and dual phase (ProtoDUNE-DP) technologies, are under construction and will be operated at the CERN Neutrino Platform (NP) starting in 2018.
\end{Abstract}
\vfill
\begin{Presented}
NuPhys2017, Prospects in Neutrino Physics

Barbican Centre, London, UK,  December 20--22, 2017
\end{Presented}
\vfill
\end{titlepage}
\def\thefootnote{\fnsymbol{footnote}}
\setcounter{footnote}{0}

\section{Introduction}
\label{sec:Introduction}
The past century has seen a revolution in neutrino physics, with neutrino oscillations first 
confirmed within the past 20 years~\cite{PhysRevLett.81.1562,PhysRevLett.87.071301,PhysRevLett.90.021802} now firmly entering the precision era.  Neutrino oscillations rely on flavour mixing in which the three neutrino flavours are a linear combination of the three neutrino mass states,
which are combined via a $3\times 3$ mixing matrix.  The matrix is typically parameterised in terms of three mixing angles ($\theta_{12},\theta_{13},\theta_{23}$) and a \gls{CP} violating phase $\delta_{\textrm{CP}}$.  Additionally, the neutrino oscillation phenomenon does not depend on the absolute mass scale but rather the neutrino mass squared difference $\Delta m_{ij}^2\equiv m_i^2-m_j^2$.
\newline
\indent
While many experiments have pushed forward our understanding of the universe by studying neutrino oscillations, some key open questions need to be addressed by future \gls{LBL} oscillation experiments.  Furthermore, a \gls{FD} in an \gls{LBL} experiment (large active mass in a low background environment) is well suited to address certain key questions beyond neutrino physics.  A broad cross section of these questions is:
\begin{itemize}
\item \textbf{\gls{CPV} in neutrino oscillations} - $\delta_{\textrm{CP}}$ is yet to be measured.  A \gls{CP} non-conversing value of $\delta_{\textrm{CP}}$ ($\delta_{\textrm{CP}}\neq0,\pi$) would be the first observation of \gls{CPV} in the lepton sector; a requirement for leptogenesis.  In some models, $\delta_{\textrm{CP}}$ can directly explain the matter-antimatter asymmetry~\cite{Pascoli:2006ci}.
\item \textbf{The neutrino \gls{MO}} - Previous experiments have only measured the magnitude of $\Delta m^2_{23}$ therefore it is not known whether the third mass state, $\nu_3$, is much heavier or much lighter than $\nu_{1,2}$.  These scenarios are respectively known as the \gls{NO} and the \gls{IO}. 
\item \textbf{The octant of $\theta_{23}$} - Previous experiments have measured degenerate values of $\theta_{23}$ that are consistent with maximal mixing ($\theta_{23}=45\degree$).  Such maximal mixing may indicate an underlying symmetry of nature.
\item \textbf{\Gls{GUT}} - \Glspl{GUT} predict baryon number violating processes which are yet to be observed such as proton decay.
\item \textbf{Core-collapse supernova} - What are the underlying mechanisms and how to neutrinos play a role?
\end{itemize}

\section{The Deep Underground Neutrino Experiment}
\Gls{DUNE}~\cite{Acciarri:2016crz, Acciarri:2015uup, Strait:2016mof, Acciarri:2016ooe} is a future 1300~km \gls{LBL} neutrino oscillation experiment, to be situated across two sites in North America and aiming to address the key open questions outlined in section~\ref{sec:Introduction}.  The near site, located at \gls{FNAL} (IL), will house the \gls{LBNF} neutrino beam and \gls{ND} complex.  At the opposing end of the baseline, the far site in \gls{SURF} (SD) will house four 10~kt \glspl{LArTPC} acting as \glspl{FD}.
\newline
\indent
\gls{LBNF}~\cite{Strait:2016mof} will provide a GeV-scale, wide-band, highly pure $\nu_\mu$ beam with an initial power of 1.2~MW, upgradeable to 2.4~MW.  Based on features of the NuMI neutrino beam~\cite{Adamson:2015dkw} at \gls{FNAL}, the target and horn system has been tuned via a genetic algorithm to maximise sensitivity to \gls{CPV}.  The beam is capable of operating in \gls{FHC} and \gls{RHC} modes which preferentially selects $\nu$ or $\bar{\nu}$ via reversal of the magnetic field in the focusing horns.
\newline
\indent
The role of the \gls{ND} is to constrain the unoscillated neutrino interaction rate in the \gls{FD}, reducing systematic uncertainties in the neutrino oscillation measurements. The \gls{ND} will also provide a rich neutrino interaction cross-section measurement program.  The \gls{ND}, which will be housed 574~m from the \gls{LBNF} target and $\sim$60~m underground, will be a composite system and is under design, consisting of a \gls{LArTPC} along with a \gls{GAr} time projection chamber or a fine-grained straw tube tracker.
\begin{figure}
\centering
\includegraphics[width=0.5\textwidth]{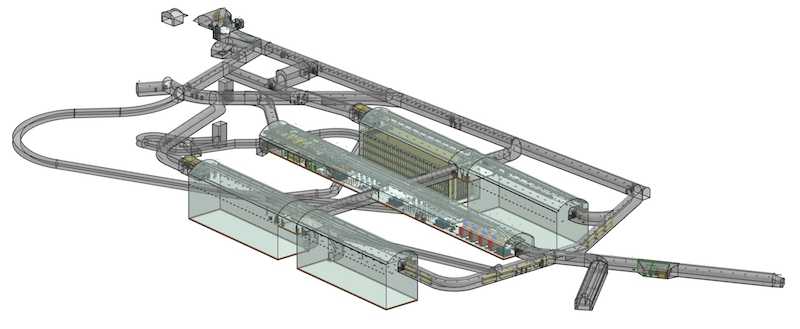}
\caption{Schematic of the \gls{FD} complex located on the 4850 level at \gls{SURF}.}
\label{fig:far-detector-configuration-oct2016}
\end{figure}
\newline
\indent
The \gls{FD} complex, located 4850~ft underground at \gls{SURF} and shown in Fig.~\ref{fig:far-detector-configuration-oct2016}, will house four 10~kt \glspl{LArTPC}~\cite{Acciarri:2016ooe}.  Two technologies are available for the \gls{FD} modules; a reference design using \gls{SP} technology and an alternative \gls{DP} design.  The \gls{SP} design consists of four horizontal drift regions per \gls{FD} module flanked by opposing anodes and cathodes which maintain a 500~V/cm electric field.  The anode is instrumented for ionisation electron readout and consists of three planes of wires.  
Conversely, the \gls{DP} design consists of a single, vertical drift volume per \gls{FD} module with a cathode on the module floor and an instrumented anode on the ceiling, maintaining a 500~V/cm electric field.  Sandwiched between the \gls{LAr} surface and the anode is a \gls{GAr} phase.  The ionisation electrons are extracted from the \gls{LAr} phase into the \gls{GAr} phase and amplified by large electron multipliers.  The amplified signal is then readout by the 2D segmented anode.
\newline
\indent
A staged \gls{FD} installation approach will be adopted allowing an early physics program.  The \gls{FD} cavern groundbreaking occurred in 2017 with installation of the first \gls{FD} module planned for 2021.  First data is expected in 2024 followed by \gls{LBNF} neutrino beam and \gls{ND} availability in 2026.  The technology used for each \gls{FD} module is dependent upon the success of the ProtoDUNE program (see section~\ref{sec:ProtoDUNE}) but both technologies are expected to be utilised.
\section{Physics reach}
\label{sec:physics_reach}
\begin{figure}
\centering
    \centering
    \subfloat[$\nu_\mu$ \gls{CC} sample in \gls{FHC} mode.]{{\label{fig:NuMuDis_no_2017}\includegraphics[width=0.45\textwidth]{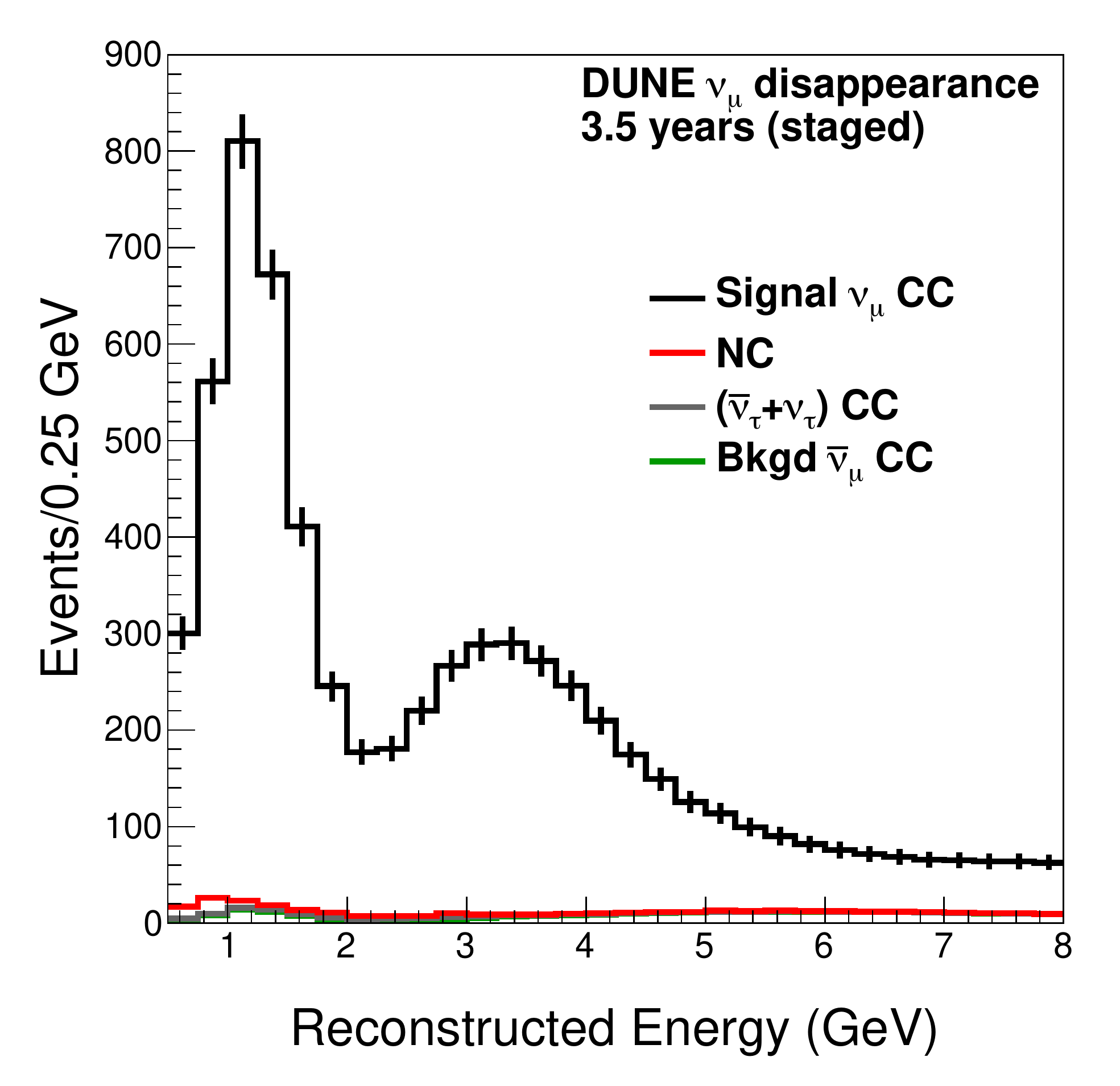} }}%
    \hspace{0.05\textwidth}
    \subfloat[$\nu_e$ \gls{CC} sample in \gls{FHC} mode.]{{\label{fig:NueApp_no_2017}\includegraphics[width=0.45\textwidth]{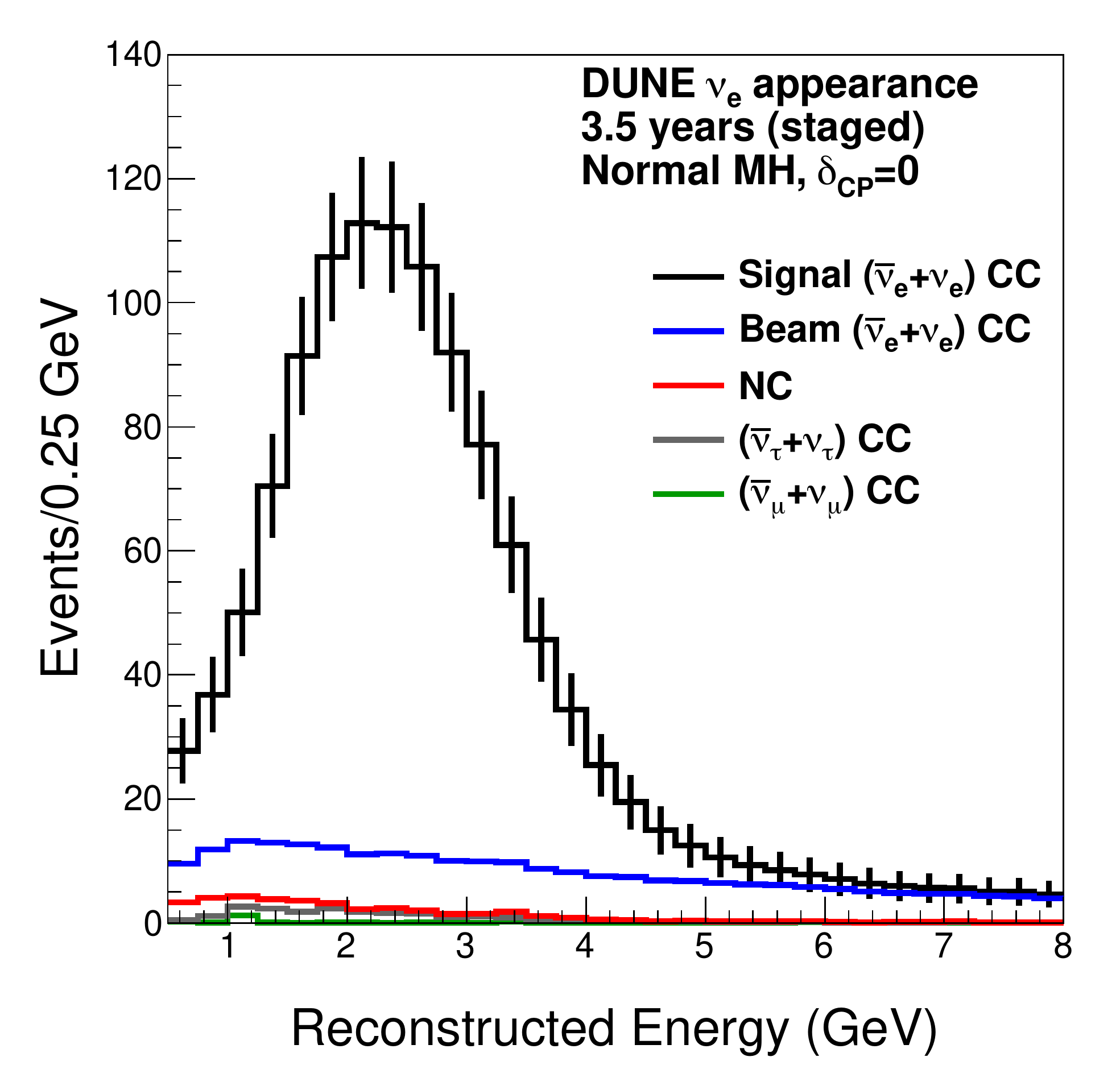} }}\\ 
    \subfloat[$\bar{\nu}_\mu$ \gls{CC} sample in \gls{RHC} mode.]{{\label{fig:NuMuBarDis_no_2017}\includegraphics[width=0.45\textwidth]{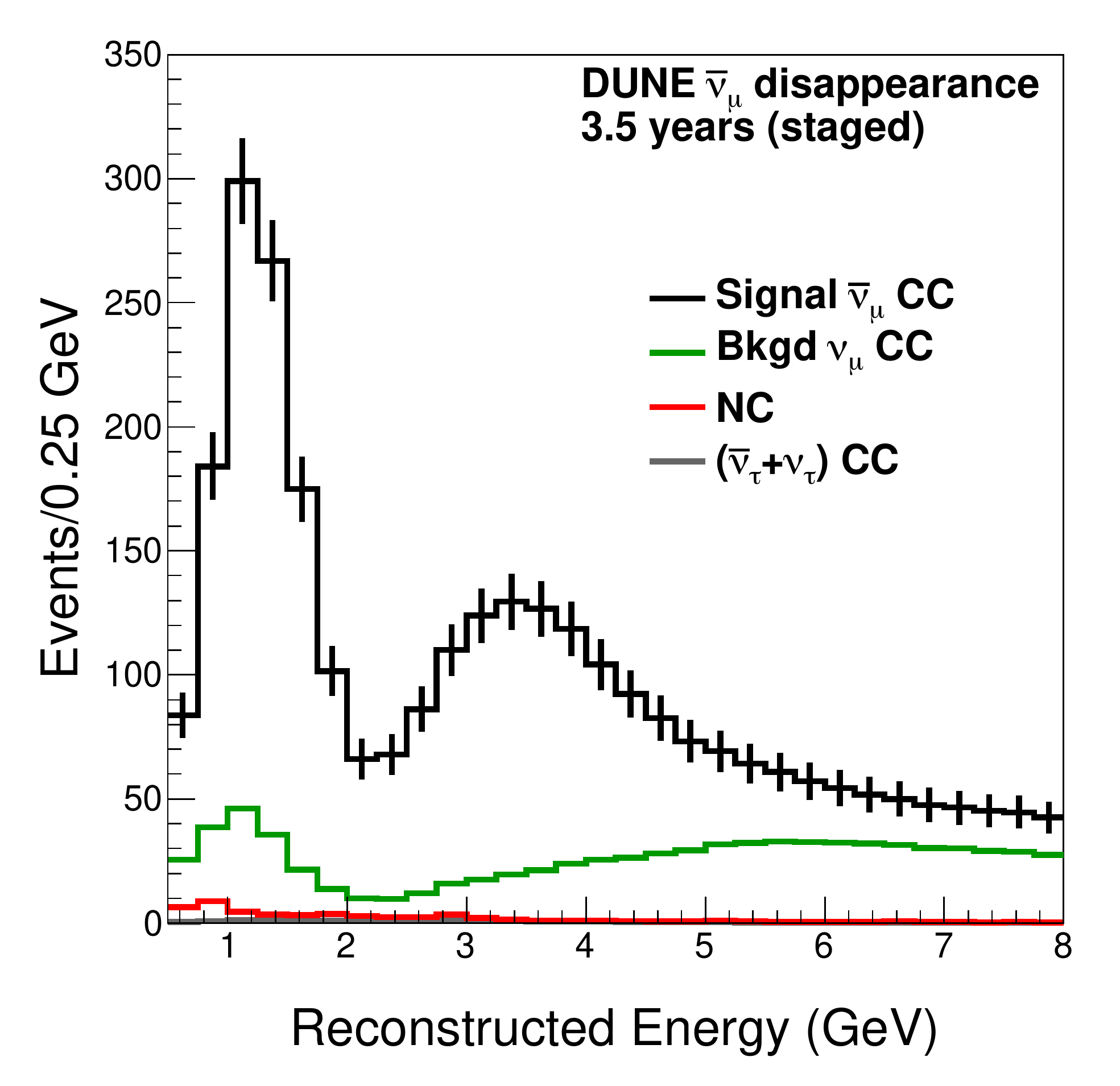} }}%
    \hspace{0.05\textwidth}
    \subfloat[$\bar{\nu}_e$ \gls{CC} sample in \gls{RHC} mode.]{{\label{fig:NueBarApp_no_2017}\includegraphics[width=0.45\textwidth]{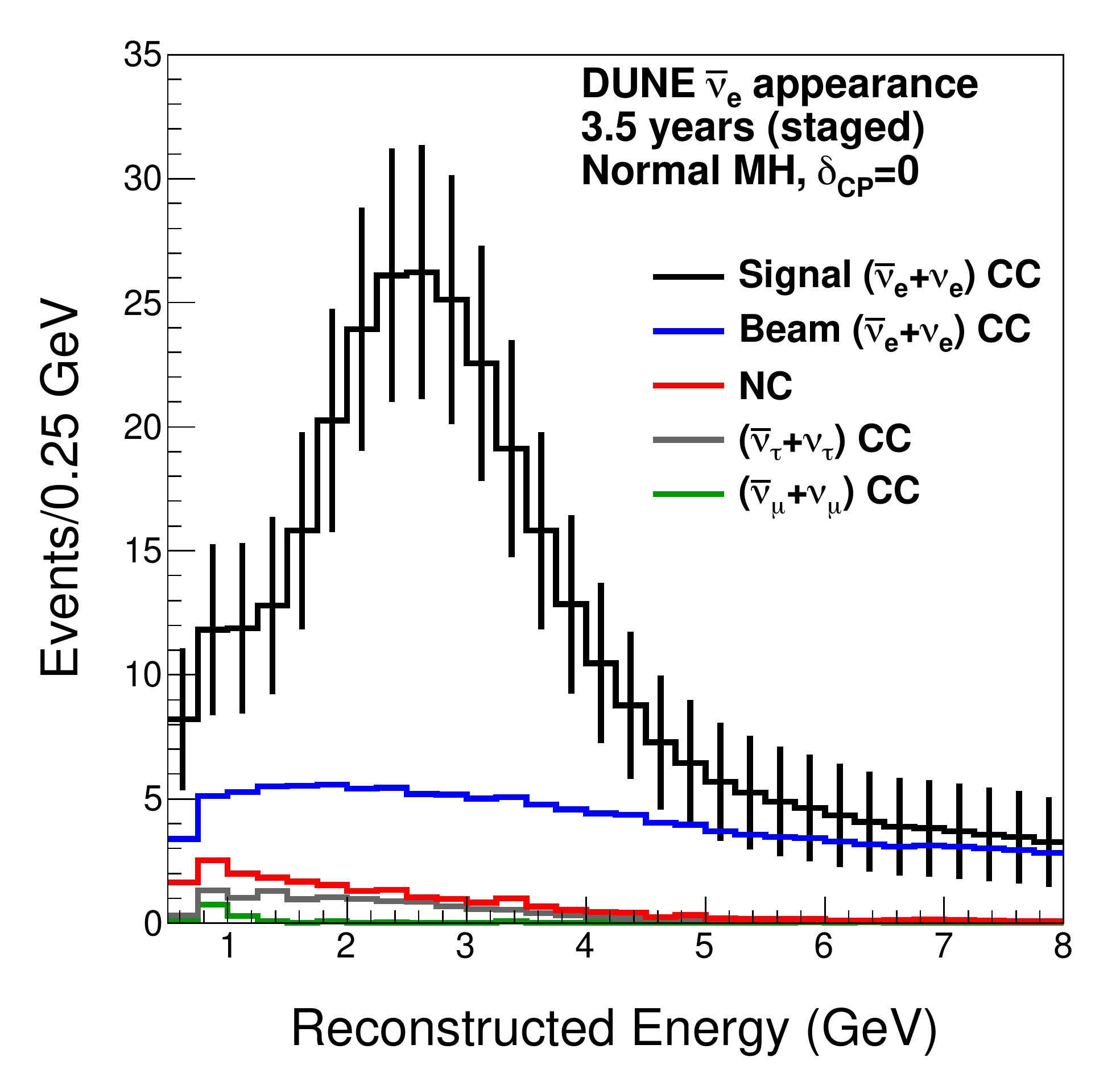} }}%
\caption{Reconstructed spectra of selected (anti-)neutrino \gls{CC} interactions showing signal (black), \gls{NC} (red), $\nu_\tau$ and $\bar{\nu}_\tau$ (grey), $\nu_\mu$ or $\bar{\nu}_\mu$ (green) and \gls{LBNF} beam intrinsic $\nu_e$ and $\bar{\nu}_e$ (blue) interactions.  Each distribution assumes a 3.5 year exposure, assuming the staging scenario described in table~\ref{table:dune_staging_scenario}. The neutrino oscillation parameters are based on NuFit~2016~\cite{Esteban2017} with $\delta_{\textrm{CP}}=0$ and true \gls{NO}.}
\label{fig:dune_fd_samples}
\end{figure}

\begin{figure}%
    \centering
    \subfloat[True \gls{NO}.]{{\label{fig:cpv_two_exps_th23band_no_2017}\includegraphics[width=0.4\textwidth]{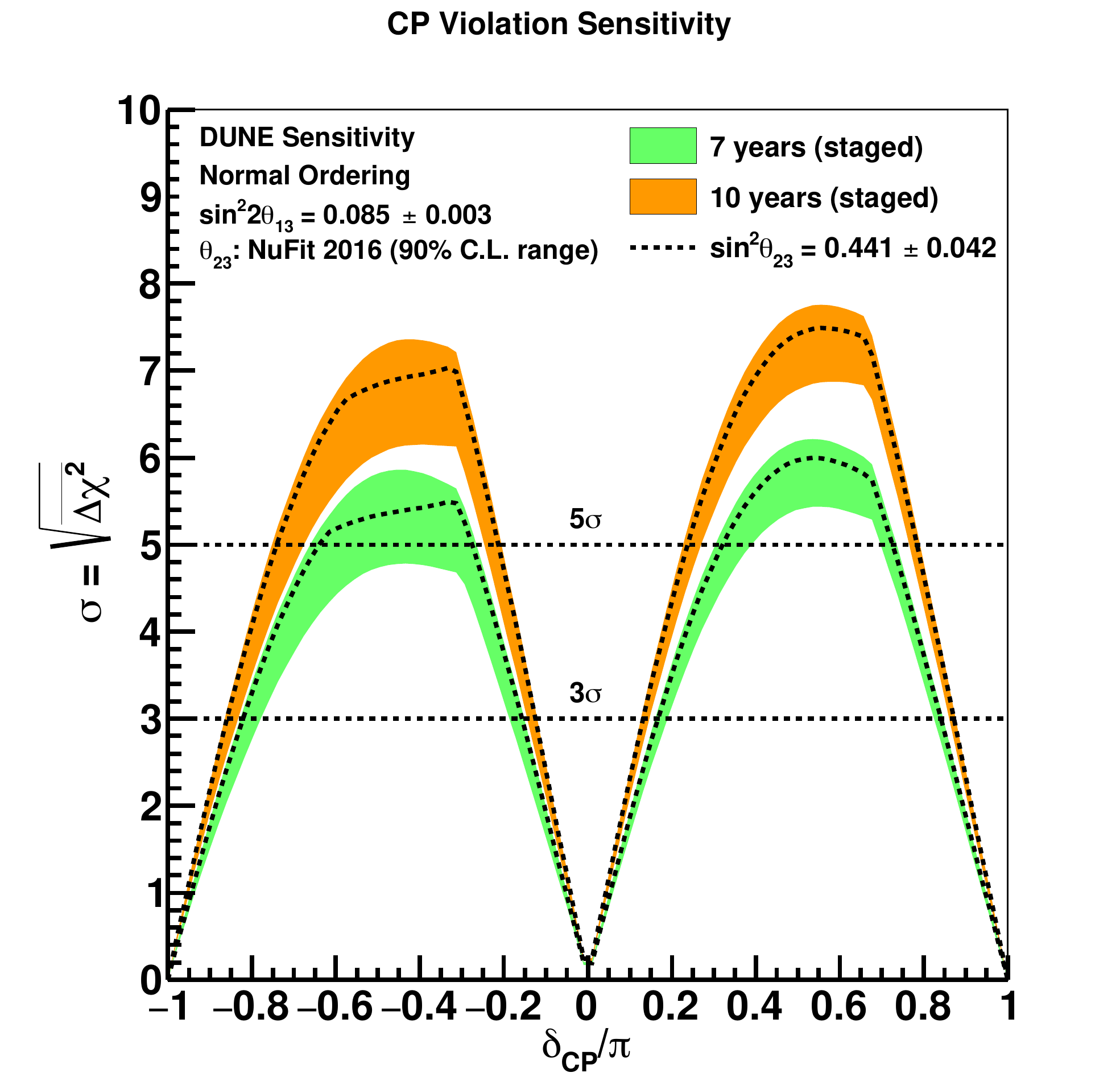} }}%
    \hspace{0.1\textwidth}
    \subfloat[True \gls{IO}.]{{\label{fig:cpv_two_exps_th23band_io_2017}\includegraphics[width=0.4\textwidth]{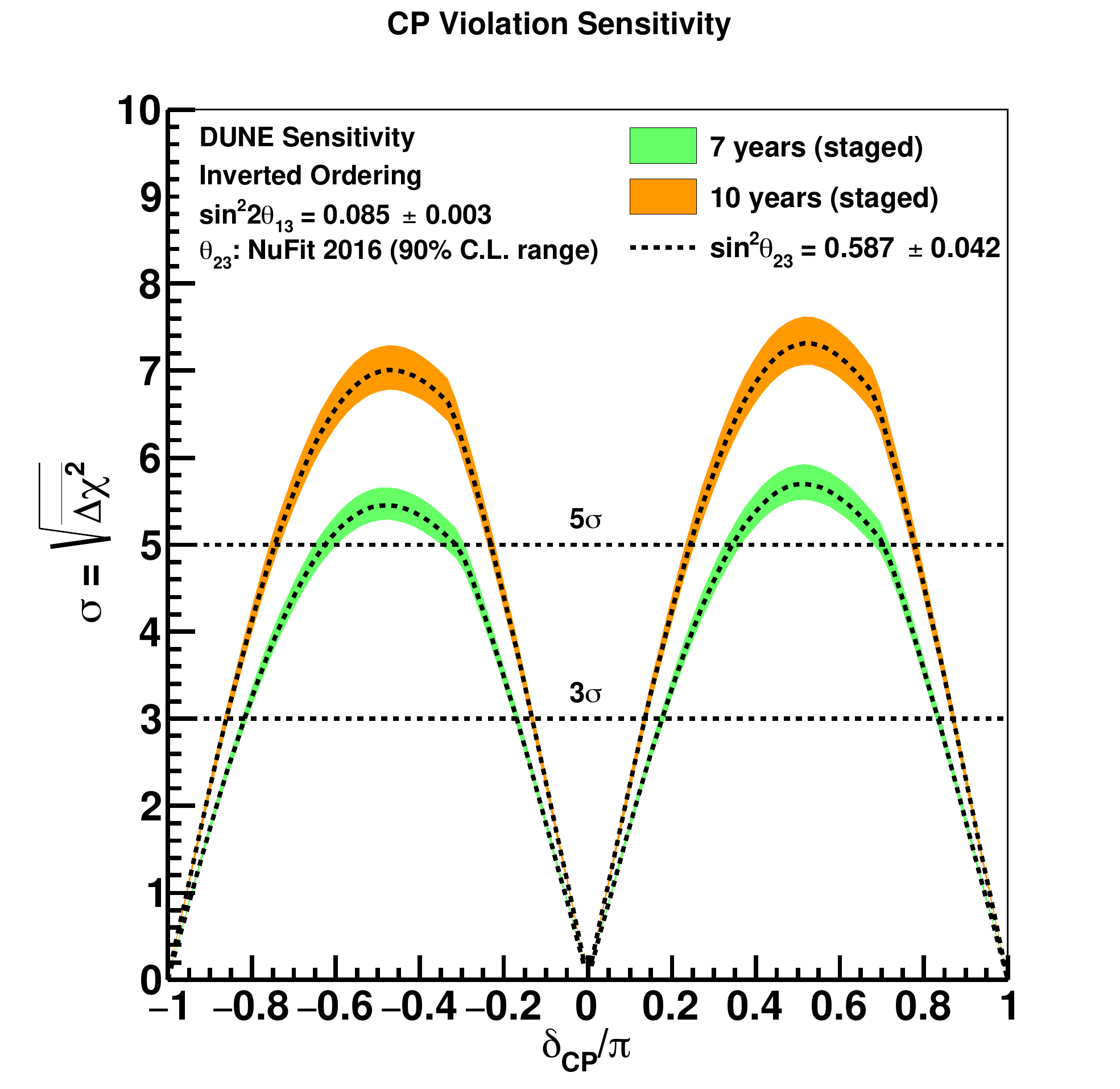} }}%
    \caption{The significance with which $\delta_{\textrm{CP}}\neq0,\pi$, can be determined as a function of the true value of $\delta_{\textrm{CP}}$, for exposures of seven (green) or 10 (orange) years, assuming the staging scenario described in table~\ref{table:dune_staging_scenario}.  The inputs to the sensitivities are shown in Fig.~\ref{fig:dune_fd_samples}.  The neutrino oscillation parameters are based on NuFit~2016~\cite{Esteban2017}.  The dashed line is the sensitivity for the NuFit~2016 central value of $\theta_{23}$ with the width of the band representing the range of sensitivities for the 90\% C.L. range of $\theta_{23}$ values.}%
    \label{fig:cpv_two_exps_th23band_2017}%
\end{figure}

\begin{figure}%
    \centering
    \subfloat[True \gls{NO}.]{{\label{fig:mh_two_exps_th23band_no_2017}\includegraphics[width=0.40\textwidth]{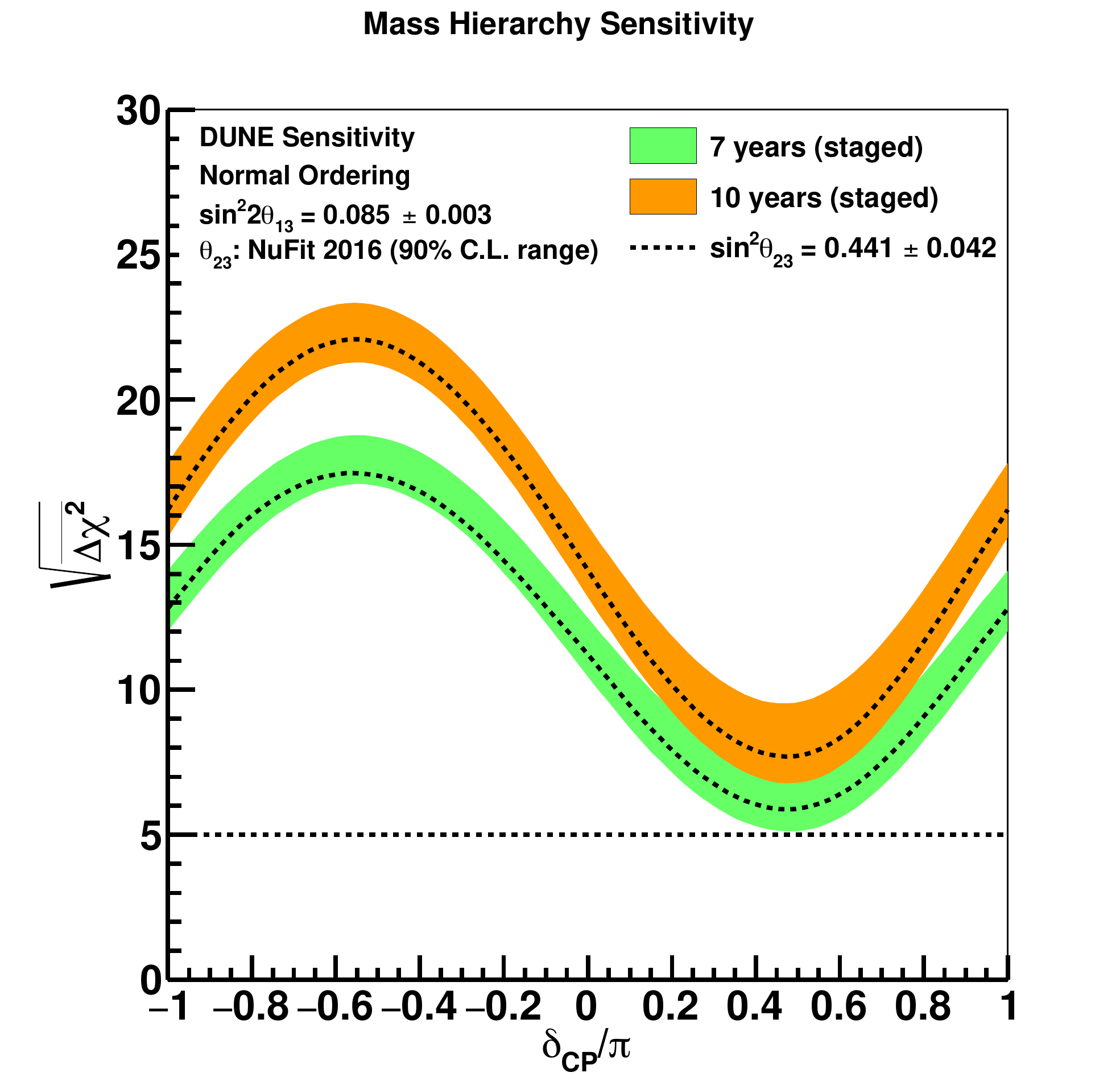} }}%
    \hspace{0.1\textwidth}
    \subfloat[True \gls{IO}.]{{\label{fig:mh_two_exps_th23band_io_2017}\includegraphics[width=0.40\textwidth]{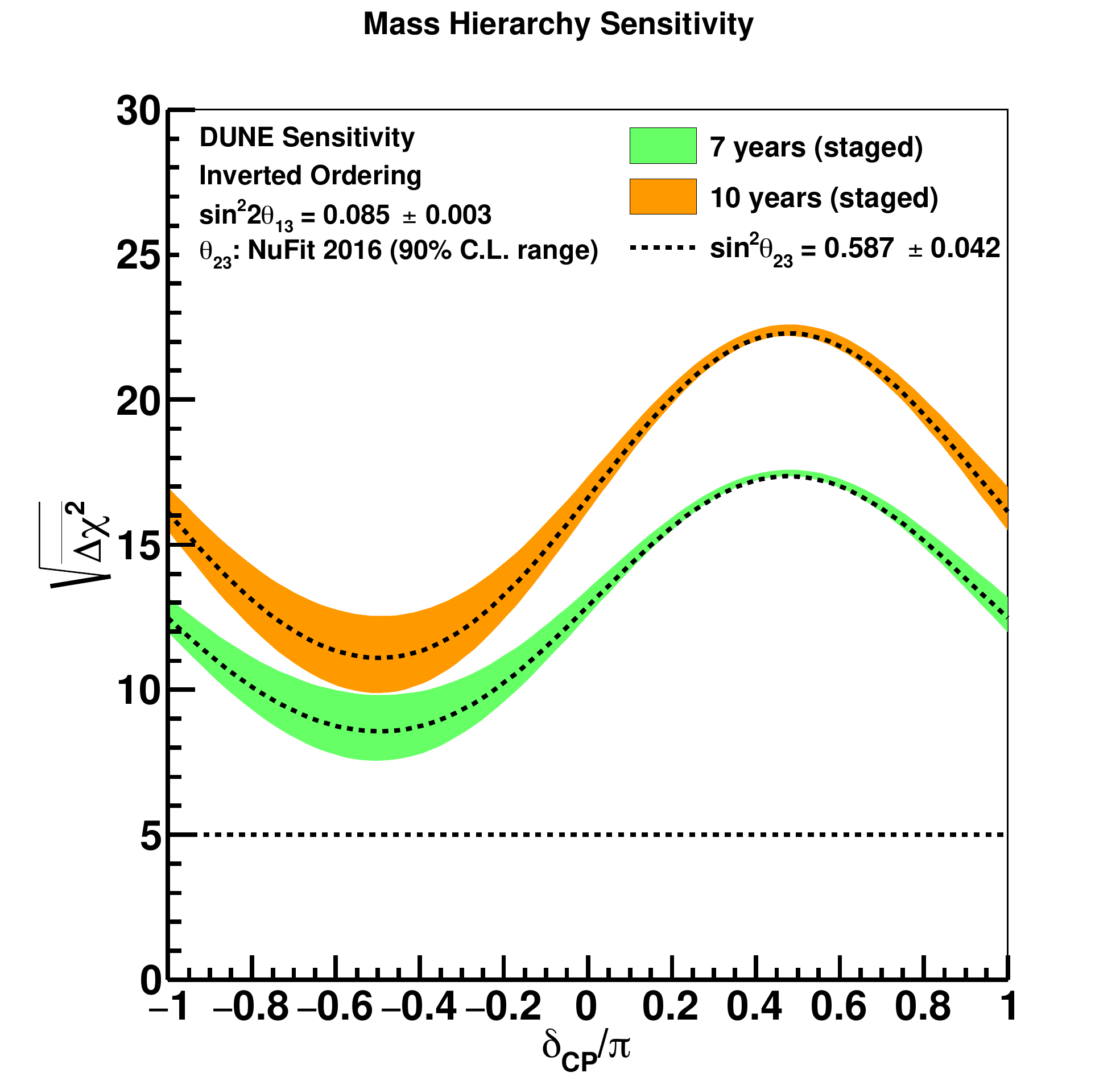} }}%
    \caption{The significance with which the neutrino \gls{MO} can be determined as a function of the true value of $\delta_{\textrm{CP}}$, for exposures of seven (green) or 10 (orange) years, assuming the staging scenario described in table~\ref{table:dune_staging_scenario}.  The inputs to the sensitivities are shown in Fig.~\ref{fig:dune_fd_samples}.  The neutrino oscillation parameters are based on NuFit~2016~\cite{Esteban2017}.  The dashed line is the sensitivity for the NuFit~2016 central value of $\theta_{23}$ with the width of the band representing the range of sensitivities for the 90\% C.L. range of $\theta_{23}$ values.}%
    \label{fig:mh_two_exps_th23band_2017}%
\end{figure}

\begin{figure}
\centering
\includegraphics[width=0.40\textwidth]{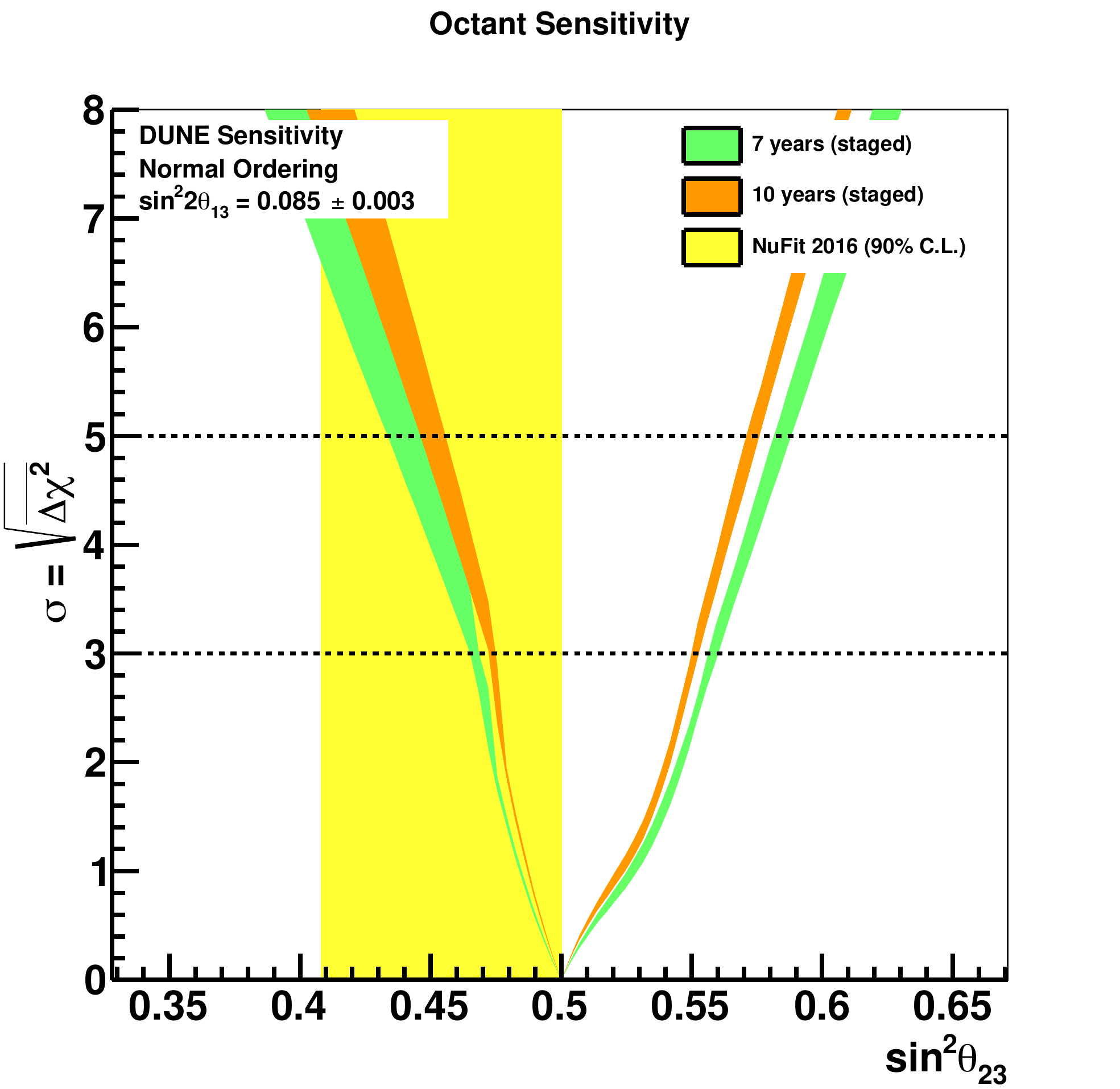}
\caption{The significance with which the $\theta_{23}$ octant can be resolved as a function of the true value of $\sin^2\theta_{23}$, for exposures of seven (green) or 10 (orange) years, assuming the staging scenario described in table~\ref{table:dune_staging_scenario}.  The inputs to the sensitivities are shown in Fig.~\ref{fig:dune_fd_samples}.  The neutrino oscillation parameters are based on NuFit~2016~\cite{Esteban2017}.  The width of the band represents the range in sensitivity due to variation in the true value of $\delta_{\textrm{CP}}$, where the band covers the top 80\% least extreme sensitivities.  The yellow shaded region indicates the NuFit 2016 $\sin^2\theta_{23}$ 90\% C.L. allowed region.}
\label{fig:octant_no_2017}
\end{figure}

\begin{figure}
\centering
\includegraphics[width=0.5\textwidth]{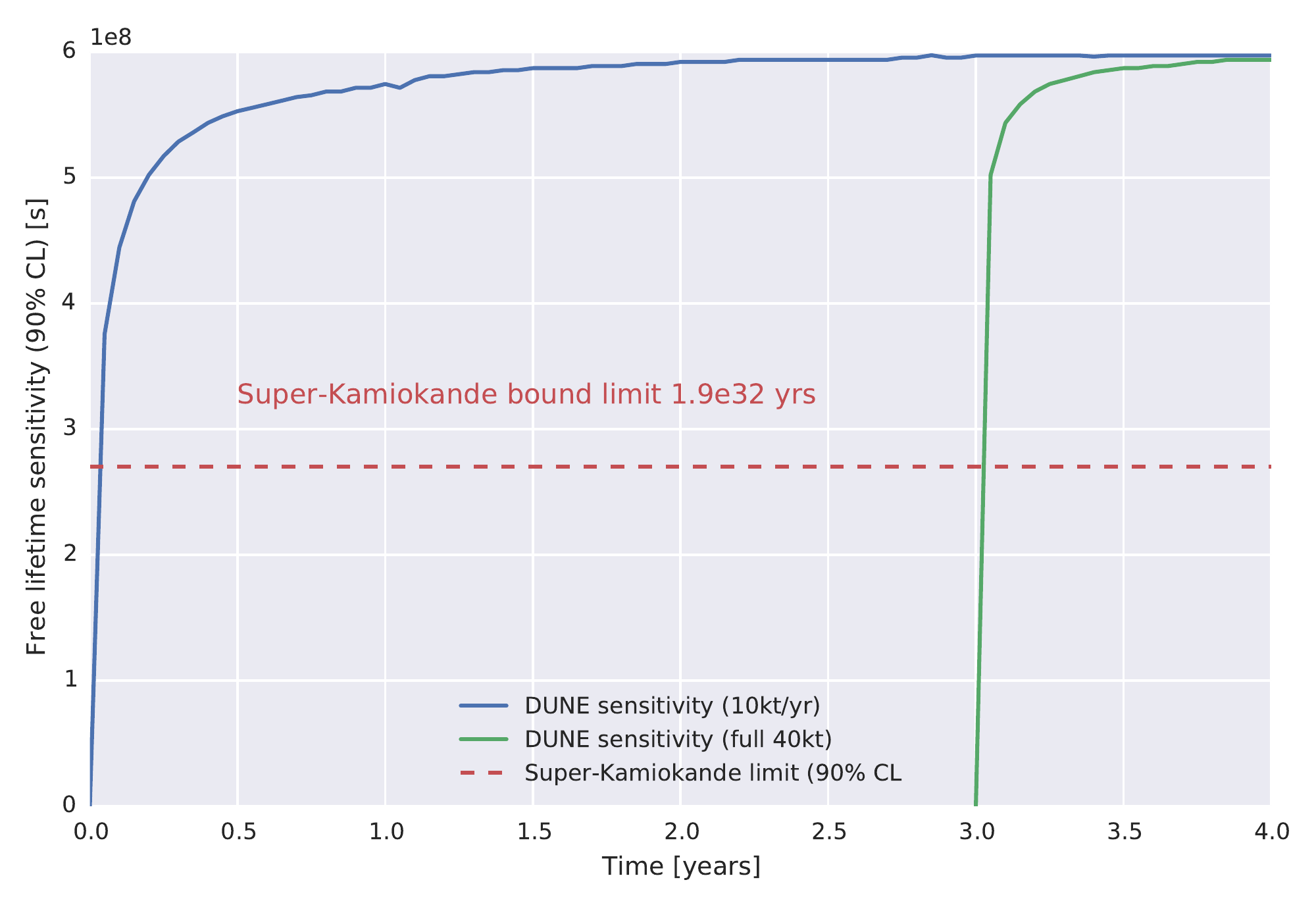}
\caption{Free $n$~---~$\bar{n}$ lifetime sensitivity as a function of years of exposure with sensitivities for two construction scenarios (solid blue and solid green lines) and compared with the Super-K limit (orange dashed line).}
\label{fig:free_sens_staged}
\end{figure}

\Gls{CPV} and the neutrino \gls{MO} introduce an asymmetry between the neutrino and anti-neutrino oscillation probabilities which causes a measurement degeneracy.  The magnitude of the \gls{MO}-induced asymmetry increase with baseline length whereas the \gls{CPV}-induced asymmetry is constant~\cite{Bass:2013vcg}.  This degeneracy can be resolved for baselines longer than 1200~km which implies that \gls{DUNE}, with a 1300~km baseline, can unambiguously determine the \gls{MO} and measure $\delta_{\textrm{CP}}$. 
\newline
\indent
The primary neutrino oscillation physics program of \gls{DUNE} relies not only on measuring the oscillated energy spectrum of multiple samples of neutrino interactions in the \gls{FD} but also measuring subtle differences between them.  To establish the physics reach of \gls{DUNE} prior to data-taking, a parameterised \gls{MC} simulation of the \gls{LBL} physics program with GLoBES~\cite{Huber:2004ka,Huber:2007ji} has been used to explore the key sensitivities of the experiment.  
This exploration is based on a staged construction scenario which assumes a 50:50 time split between \gls{FHC} and \gls{RHC} \gls{LBNF} beam running and is summarised in table~\ref{table:dune_staging_scenario}. 
\begin{table}
\begin{center}
\begin{tabular}{ccccc}  
Year &  \thead{Number of installed\\ \gls{FD} modules} &  \thead{Total mass of installed\\ \gls{FD} modules (kt) } &  
\thead{\Gls{LBNF} beam\\ power (MW)} &\thead{Exposure at\\ year end (kt~MW~yr) } \\ \hline
1 & 2 & 20 & 1.07 & 21 \\
2 & 3 & 30 & 1.07 & 54 \\
4 & 4 & 40 & 1.07 & 128 \\
7 & 4 & 40 & 1.07 & 300 \\
10 & 4 & 40 & 2.14 & 556 \\
\end{tabular}
\caption{\Gls{DUNE} staged data-taking scenario.  The scenario assumes a 50:50 time split between \gls{FHC} and \gls{RHC} \gls{LBNF} beam running modes.}
\label{table:dune_staging_scenario}
\end{center}
\end{table}
\newline
\indent
The parameterised \gls{MC} (the Fast Monte Carlo simulation~\cite{Acciarri:2016crz}) outputs selected \gls{CC} reconstructed spectra which are shown in Fig.~\ref{fig:dune_fd_samples} 
with applied oscillations using the NuFit 2016~\cite{Esteban2017} parameters.  Each distribution shows what level of statistics would be expected after 3.5~years of data-taking in separate \gls{FHC} and \gls{RHC} \gls{LBNF} beam running modes (seven~years total).
\newline
\indent
The sensitivities to both lepton sector \gls{CPV} and the neutrino \gls{MO} are obtained from a fit to all four oscillated neutrino samples detailed in Fig.~\ref{fig:dune_fd_samples}.  
The significance with which \gls{DUNE} can observe lepton sector \gls{CPV} ($\delta_{\textrm{CP}}\neq0,\pi$) is shown in Fig.~\ref{fig:cpv_two_exps_th23band_2017} as a function of the true value of $\delta_{\textrm{CP}}$ for true \gls{NO} and \gls{IO}.  Partial $5\upsigma$ coverage occurs after seven~years of data-taking with the coverage dramatically increasing as exposure increases to 10~years.
\newline
\indent
The significance with which \gls{DUNE} can determine the neutrino \gls{MO} as a function of the true value of $\delta_{\textrm{CP}}$ is shown in Fig.~\ref{fig:mh_two_exps_th23band_2017} for both true \gls{NO} and \gls{IO}.  Driven by the 1300~km baseline, Fig.~\ref{fig:mh_two_exps_th23band_2017} shows greater than $5\upsigma$ sensitivity to the neutrino \gls{MO} for all possible values of $\delta_{\textrm{CP}}$.
\newline
\indent
Unlike \gls{CPV} and the neutrino \gls{MO}, establishing the octant of $\theta_{23}$ relies only a combined measurement of the $\nu_\mu\rightarrow\nu_\mu$ (which is sensitive to $\sin^{2}2\theta_{23}$) and $\nu_\mu\rightarrow\nu_e$ (which is sensitive to $\sin^{2}\theta_{23}$) oscillation channels.  
The significance with which \gls{DUNE} can establish the octant of $\theta_{23}$ as a function of the true value of $\sin^{2}\theta_{23}$ is shown in Fig.~\ref{fig:octant_no_2017}.  
Approximately 50\% of the NuFit~2016 $\sin^2\theta_{23}$ C.L. region is covered with $5\upsigma$ signifance after 10~years of data-taking.
\newline
\indent
The scope of \gls{DUNE} extends into other \gls{BSM} physics.  \Glspl{GUT} predict proton lifetimes possibly within reach of the full \gls{DUNE} experiment.  
The $p\rightarrow K^{+}\bar{\nu}$ decay channel is uniquely suited to a \gls{LArTPC} due to the high ionization density of the decay $K^{+}$ which results in high identification efficiency.
\newline
\indent
\Gls{DUNE} is also sensitive to $n$~---~$\bar{n}$ oscillations in which a nucleus-bound neutron converts to an anti-neutron that annihilates with another bound nucleon.  This results in an isotropic burst of neutral and charged pions which resembles a distinctive star-like topology.  A free lifetime sensitivity has been established using a \gls{CNN} to select this topology over the dominant atmospheric neutrino interaction background which is shown in Fig.~\ref{fig:free_sens_staged}.

\section{ProtoDUNE}
\label{sec:ProtoDUNE}
The \gls{DUNE} \glspl{FD}, with a combined 40~kt \gls{LAr} mass, will be the largest \glspl{LArTPC} ever constructed.  This demands extensive R\&D to establish the production process, understand subsystem integration and to validate that the design and performance is sufficient for the core physics program.  To achieve this goal, two large scale prototype \glspl{LArTPC}, named ProtoDUNE-SP~\cite{Abi:2017aow} and ProtoDUNE-DP, which will use the complementary \gls{SP} and \gls{DP} technologies, are currently under construction at CERN and will be exposed to a GeV-scale charged particle test beam.  Both prototypes will each hold 770~t of \gls{LAr} and will use full scale \gls{FD} components.  
Extensions of the H2 and H4 SPS beamlines will service ProtoDUNE-DP and ProtoDUNE-SP respectively.
\newline
\indent
The overarching aim of ProtoDUNE measurement program is acquiring a better understanding of interaction processes on argon and to optimise \gls{LArTPC} reconstruction techniques.  
Charged pions and protons from the beam will be used to study hadronic interaction mechanisms.  The electron component will be used to further develop \gls{LArTPC} shower reconstruction algorithms in addition to expanding photon/electron separation techniques.  Interactions from GeV-scale charged pions will produce as vast quantity of charged kaons, allowing for study of charged kaon reconstruction and identification efficiencies.  
Muons will provide a standard candle for dE/dx calibration with Michel electrons from stopping muons providing a low energy benchmark for supernova neutrino interactions.
\newline
\indent
The ProtoDUNE cryostats are complete and are housed in the EHN1 building at the CERN neutrino platform. The final full-scale sub-components of the \glspl{LArTPC} are now being shipped from various construction sites in North America and Europe to CERN where installation is taking place.  Commissioning and data-taking is expected to take place in 2018 prior to the SPS long shutdown.

\section{Conclusions}
The neutrino oscillation field is progressing rapidly but some key open questions remain.  \gls{DUNE}, which will utilise \gls{LArTPC} technology, will be able to unambiguously determine the neutrino \gls{MO} and probe lepton sector \gls{CPV} in a timely manner.  The extended physics program will further probe neutrino oscillations and \gls{BSM} physics.
\newline
\indent
Two kt scale \glspl{LArTPC} are currently under construction at CERN (the ProtoDUNE-SP and ProtoDUNE-DP prototypes) that will critically test the engineering and construction procedures of the proposed \gls{FD} technologies whilst providing a rich physics program of direct use in the key measurements of \gls{DUNE}.

\bibliographystyle{h-physrev}
\bibliography{bibliography.bib}

\end{document}